# Demonstration and operation of quantum harmonic oscillators in an AlGaAs-GaAs heterostructure

Guangqiang Mei, Pengfei Suo, Li Mao, Min Feng, Limin Cao[*]

*School of Physics and Technology, Center for Nanoscience and Nanotechnology, and Key Laboratory of Artificial Micro- and Nano-structures of Ministry of Education, Wuhan University, Wuhan 430072, China*

Corresponding author. E-mail: [*]limincao@whu.edu.cn

The quantum harmonic oscillator (QHO), one of the most important and ubiquitous model systems in quantum mechanics, features equally spaced energy levels or eigenstates. Here we present a new class of nearly ideal QHOs formed by hydrogenic substitutional dopants in an AlGaAs/GaAs heterostructure. On the basis of model calculations, we demonstrate that, when a δ-doping Si donor substitutes the Ga/Al lattice site close to AlGaAs/GaAs heterointerface, a hydrogenic Si QHO, characterized by a restoring Coulomb force producing square law harmonic potential, is formed. This gives rise to QHO states with energy spacing of ~8-9 meV. We experimentally confirm this proposal by utilizing gate tuning and measuring QHO states using an aluminum single-electron transistor (SET). A sharp and fast oscillation with period of ~7-8 mV appears in addition to the regular Coulomb blockade (CB) oscillation with much larger period, for positive gate biases above 0.5 V. The observation of fast oscillation and its behavior is quantitatively consistent with our theoretical result, manifesting the harmonic motion of electrons from the QHO. Our results might establish a general principle to design, construct and manipulate QHOs in semiconductor heterostructures, opening future possibilities for their quantum applications.

# 1 Introduction

The quantum harmonic oscillators (QHOs) are central to describe the most elementary structure of quantum world, ranging from trapped electrons, atoms and/or quantum particles around the minimum of a potential well, vibrations in molecules and phonons in solids, to bosons and quantum field theory. The robust generation and manipulation of QHO states may facilitate the studies and explorations of quantum sensing, quantum simulations, quantum information processing/storage, etc., a reality at large scales [1-19]. As an example, the ion-trap qubit architecture built using the electronic states of trapped ion oscillators has been established as one of the most promising systems for a scalable, universal quantum computer [2-6]. To date, artificial QHOs formed by ions and atoms in electromagnetic traps and quantized mechanical modes of a macroscopic solid have been demonstrated [2-16]. However, the large scale construction, reliable manipulation, high precision detection of artificial QHOs in semiconductors are limited and challenging, especially in practical applications where the existing industry-standard materials and semiconductor fabrication principles are essentially required.

In this work, combining model calculations, band engineering of semiconductor heterostructures, gate tuning, and SET device fabrication and measurements, we demonstrate that the hydrogen-like QHOs and their integration with other quantum devices (e.g., SETs) can be realized using industry-accepted semiconductor materials and technologies. Our approach is applicable in a variety of semiconductor systems for the scalable construction and reliable operation and detection of artificial QHOs.

# 2 Results and discussion

We start from the design proposal of band engineering with the well-known and widely used AlGaAs/GaAs heterostructure. The crystal and band structures of a typical AlGaAs/GaAs heterostructure are shown schematically in Fig. S1, Supplementary Information. As a natural result of band engineering, a triangular quantum potential well forms at the atomic heterointerface of AlGaAs/GaAs, where free electrons are trapped and accommodated to form the so-called two dimensional electron gas (2DEG).

Si δ-doping, which is separated from the well by a thin layer of intrinsic AlGaAs, is commonly used to modulate the carrier density of 2DEG. We then closely scrutinize the single substitutional Si atom which replaces an Al/Ga host in the lattice close to AlGaAs/GaAs heterointerface. Because Si has one more valence electron than that of Al/Ga, it functions as a donor here, forming eventually a hydrogen-like $Si^+$ center since the released electron is confined inside the interface potential well [Fig. 1(a)]. Based on the hydrogenic atom model and effective mass approximation [20-24], the Hamiltonian for the hydrogen-like $Si^+$ with a spherically symmetric screened Coulomb potential and an electron effective mass $m^*$ is

$$\widehat{H} = -\frac{\hbar^2}{2m^*}\nabla^2 - \frac{e^2}{4\pi\varepsilon_r\varepsilon_0 r} \tag{1}$$

where $\hbar$ is the reduced Planck constant, $e$ is the elementary charge, $\varepsilon_0$ is the vacuum permittivity, $\varepsilon_r$ is the relative permittivity of the system, and $r$ is the distance of orbital electron from the nucleus. Solving the time independent Schrödinger equation $\widehat{H}\psi(\boldsymbol{r}) = E\psi(\boldsymbol{r})$, we obtain the localized ground state or lowest hydrogen-like atomic orbital that is spherically centered on $Si^+$ in real space with orbit radius (or the effective Bohr radius) $R_1$ as

$$R_1 = \frac{4\pi\varepsilon_r\varepsilon_0\hbar^2}{m^*e^2} = \varepsilon_r\frac{m_e}{m^*}R_H \tag{2}$$

where $R_H = \frac{4\pi\varepsilon_0\hbar^2}{m_e e^2} = 5.29\times 10^{-11}$ m is the Bohr radius, and $m_e$ is the electron mass. From Eq. (2), we obtain the parameters of hydrogen-like $Si^+$ as $R_1 \approx 7.70$ nm in typical $Al_{0.25}Ga_{0.75}As$ where $\varepsilon_r \approx 12.19$ and $m^* \approx 0.084 m_e$, and $R_1 \approx 7.24$ nm in typical $Al_{0.3}Ga_{0.7}As$ where $\varepsilon_r \approx 12.05$ and $m^* \approx 0.088 m_e$. It is clear that the ground state orbit radius of hydrogenic Si atom/ion in AlGaAs/GaAs is much larger than that of hydrogen atom.

We then consider the interaction of external electrostatic field with individual hydrogenic Si impurity atom/ion [Fig. 1(b) and 1(c)]. An external electromagnetic stimulus may stimulate one electron back into the lowest hydrogen-like atomic orbital of $Si^+$, forming the neutral state hydrogenic Si atom [Fig. 1(b)]. In the quantum regime, the hydrogen-like Si atom under its ground state is that the electron appears statistically in spherical volume centered on the $Si^+$ nucleus within radius $R$, forming the so-called diffuse electron cloud, as shown in the light-blue spherical volume in Fig. 1(b). Here, $R = 1.5R_1$ is the expected value of radial distance of the probability electron cloud. When the separation of neighboring Si is larger than two times the effective Bohr radius ($2R_1$), it is reasonable for them to be treated as individual hydrogen-like Si atoms/ions. Under equilibrium conditions, the massive point-like nucleus Si$^+$ is at the center of electron cloud. However, when a small displacement occurs between the nucleus and electron cloud center, as shown in Fig. 1(c), in response to the displacement the electron experiences a Coulomb force

$$\boldsymbol{F}_{ne} = -\frac{e^2 \boldsymbol{r}}{4\pi\varepsilon_r\varepsilon_0 R^3} \qquad (3)$$

where $\boldsymbol{r}$ is the displacement vector of electron cloud center away from nucleus. It is clear that the Coulomb force is proportional to the displacement when the electron is displaced slightly from its equilibrium position [Fig. 1(c)]. The Coulomb interaction gives rise to a harmonic restoring force that behaves resembling an electron being harmonically bound to the nucleus via a hypothetical spring with spring constant $k$. Comparing Eq. (3) with Hooke's law $\boldsymbol{F} = -k\boldsymbol{x}$, the hydrogen-like Si atom/ion works as a harmonic oscillator, in which the spring constant $k$ is

$$k = \frac{e^2}{4\pi\varepsilon_r\varepsilon_0 R^3} \qquad (4)$$

From Eq. (4), we obtain the resonant/natural frequency of the oscillator

$$\omega_0 = \sqrt{\frac{k}{m^*}} = \sqrt{\frac{e^2}{4\pi\varepsilon_r\varepsilon_0 m^* R^3}} \qquad (5)$$

The above model calculations further imply that the hydrogen-like atom/ion of Si donor possesses a spherically-symmetric parabolic potential scaling as $r^2$

$$V(r) = \frac{1}{2}m^*\omega_0^2 r^2 = \frac{1}{2}\frac{e^2}{4\pi\varepsilon_r\varepsilon_0 R^3} r^2 \qquad (6)$$

where $r$ is the displacement of electron cloud center away from nucleus. The square law potential is the signature of an QHO for a quantum particle. It is known that an QHO uniquely features evenly spaced energy levels with spacing of $\Delta E = \hbar\omega_0$. We then obtain, from the quantum point of view, a hydrogen-like Si QHO with equally spaced QHO states with energy spacing of

$$\Delta E = E_{n+1} - E_n = \frac{\hbar e}{\sqrt{4\pi\varepsilon_r\varepsilon_0 m^* R^3}} \qquad (7)$$

From Eq. (7), we calculate the energy level spacing is ~8.34 meV for hydrogenic Si QHO in $Al_{0.25}Ga_{0.75}As$, and ~8.98 meV in $Al_{0.3}Ga_{0.7}As$.

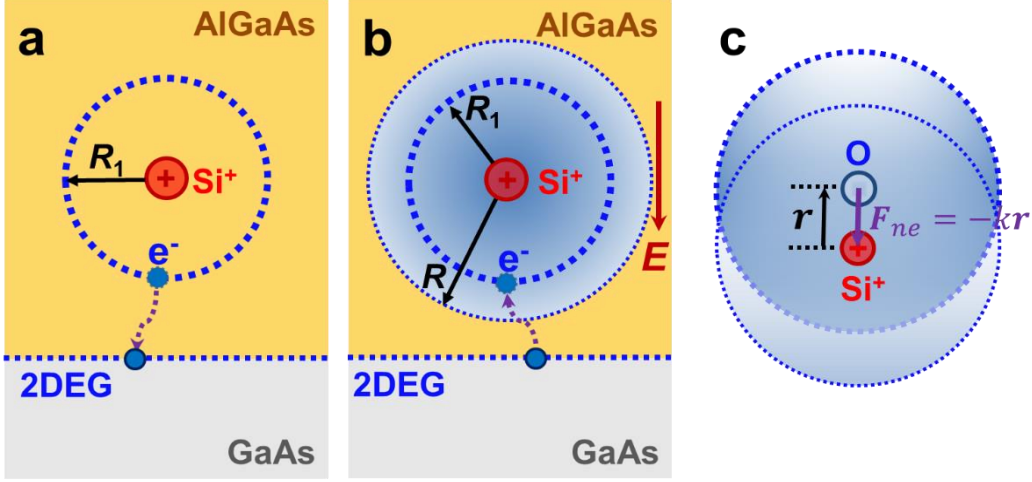

**Fig. 1** Schematics of model calculations for the formation of a hydrogenic Si QHO in AlGaAs/GaAs heterostructure. (**a**) A hydrogen-like $Si^+$ center is formed resulting from the released valence electron being trapped in the heterointerface quantum well. (**b**) Under external stimulus, such as an electric field, an electron may re-fill the lowest hydrogen-like orbit, forming the neutral hydrogen-like Si atom. Under its ground state, the electron forms the diffuse electron cloud within the spatial extent of radial radius $R = 1.5R_1$, here $R_1$ is the effective Bohr radius. (**c**) A small displacement between the electron cloud center (denoted by O) and $Si^+$ nucleus introduces a restoring Coulomb force that is proportional to displacement resembling the Hooke harmonic oscillator. The harmonic restoring force gives rise to an inherent square law potential, which is the signature of an QHO for a microscopic quantum system.

After fulfilling the model calculations, we then verify experimentally the proposed QHOs. From the discussions above, we know that the hydrogenic Si QHO in AlGaAs/GaAs possesses evenly spaced energy levels with spacing of 8-9 meV, but these energy levels are naturally unoccupied above the Fermi energy ($E_f$). These characteristics make the detection and manipulation of QHO states, however, challenging.

Our previous research on single-electron devices [25] reminds us that the SET is the most sensitive quantum electrometer [26-29]. This gives us a clue that, if we construct an integration device system where the hydrogen-like Si QHO is coupled to an SET, we may be able to effectively detect and operate the QHO in SET-QHO architecture.

For the prototype QHO-SET coupled device, we adopted the device design similar as that in Ref. 25. To verify our idea and proposal, we here only need the SET with tuning gate electrodes to couple to the underneath QHO, as shown schematically in Fig. 2.

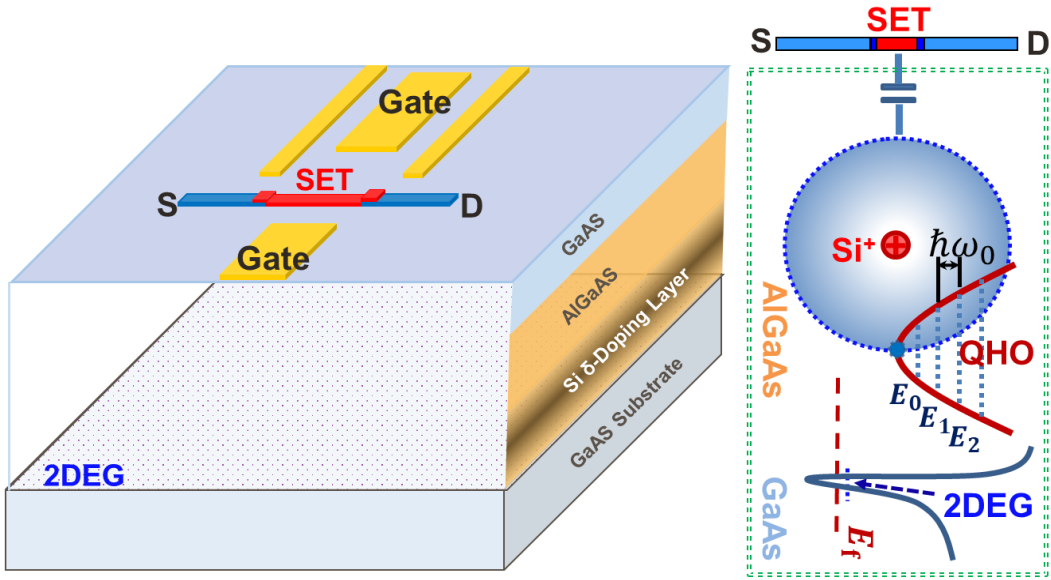

**Fig. 2** Schematic of the SET-QHO device architecture. The SET is fabricated on the surface of AlGaAs/GaAs 2DEG heterostructure substrate. The hydrogenic QHOs take place in the Si δ-doping layer. The SET is capacitively coupled to the hydrogen-like Si QHOs beneath the surface. The gate electrodes are used to tune the potentials of both SET and QHOs, and manipulate and operate them simultaneously. S, D, and Gate denote the source, drain, and gate electrodes, respectively.

The devices were fabricated using the standard two-angle shadow evaporation of aluminum on an $Al_{0.25}Ga_{0.75}As$/GaAs heterostructure (see Methods in the Supplementary Information for details of device fabrication). The concentration of the δ-doping Si directly determines the QHO distribution. Currently, we do not know the optimized substrate structure and doping concentration of Si for the formation of QHOs.

The wafer used in the experiments has 2DEG located 100 nm below the surface. The δ-doping Si dopants locate on atomic layers of about 15 nm above 2DEG, separated from 2DEG by undoped AlGaAs layer. The Si δ-doping concentration is about $2.5 \times 10^{11}$ cm$^{-2}$, corresponding to average distance of ~20 nm between two Si$^+$ ions. This means that we can reasonably treat each of the Si donors as an isolated single hydrogen-like atom/ion..

Figure 3 shows a scanning electron microscopy (SEM) image of the device fabricated on AlGaAs/GaAs heterostructure. In the device architecture, two SETs labelled as SET1 and SET2 are coupled to the hydrogen-like Si QHOs beneath the surface. For each SET, two gate electrodes marked as G$_{1(2)T}$ and G$_{1(2)B}$ serve as the top and bottom gates, respectively, and are used to tune the potentials of SET islands. We will show below that the two gates are also used to manipulate QHOs beneath the surface simultaneously. For an SET, a capacitively coupled gate can periodically tune its coulomb potentials, resulting in a periodic conductance oscillation called Coulomb blockade (CB) oscillation with the periodicity of $\Delta V_g = e/C_g$. Here, $V_g$ is the gate bias voltage applied, $C_g$ is the gate-SET capacitance. It is important to stress that for an SET, the periodic CB conductance peaks take place uniformly in both of the positive and negative gate bias regions, corresponding to the addition or removal of electrons one by one in SET island. In the following experiment we will show that, for positive biases, the SET will record changes in QHO-SET electromagnetic environment that are not correlated with CB peaks due to the addition of electrons in SET island.

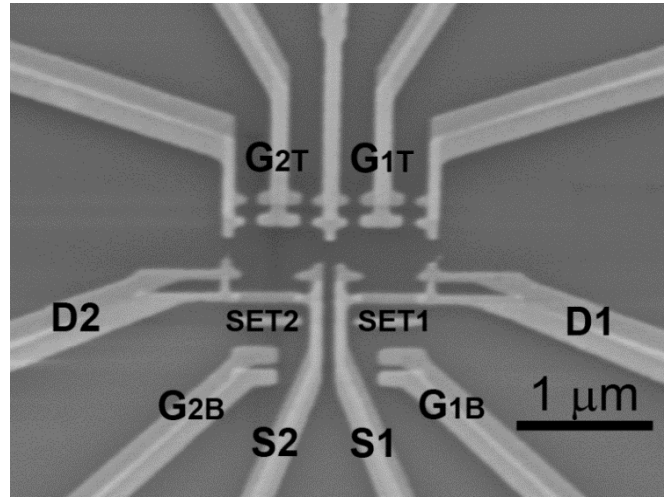

**Fig. 3** SEM image of the typical device design for the detection and operation of QHOs. D1(2) and S1(2) denote the drain and source terminals of SET1(2), and $G_{1T(B)}$ and $G_{2T(B)}$ denote the top and bottom gates of SET1(2), respectively.

The hydrogen-like Si QHO is confined in the AlGaAs/GaAs heterostructure, therefore it is impossible to characterize directly. Also, the discrete set of QHO states possesses energy above the localized lowest hydrogen-like atomic orbital: this implies that they are naturally unoccupied by electrons. To inject an electron into the originally empty electron states, equivalent to classically bonding a mass to the Hooke spring, is the prerequisite to characterize an QHO. To do so, we manage to take advantage of the gate-induced electrostatic field to tune the energy levels. It is known that, for the gate tuning, a positive gate bias induced electrostatic potential drives the bound electron states to shift to lower energies. We then conclude that a positive gate voltage needs to be applied on top of the QHO to drive its energy levels lower close to $E_f$ to acquire an electron.

The experiments of electrical measurements were carried out in a $He^3$ refrigerator operating at a base temperature of ~300 mK (see Methods in the Supplemantary Information for details of device measurements). After confirming that there is no leakage between all of the metal gate electrodes and 2DEG, and the SETs have good

Coulomb blockade performance (Fig. S2, Supplementary Information), we performed the electrical measurements using SETs and tuning gates to detect and manipulate QHOs.

Figure 4 displays the current passing through SET1 ($I_{DS1}$) as a function of the gate bias voltage applied to top gate $G_{1T}$. As predicted, $I_{DS1}$ demonstrates dramatically different features in negative bias region compared with those in positive region, as shown in Fig. 4(a). In negative bias region, the SET shows a few conductance peaks with spacing of ~440 mV, giving a nominal gate-SET capacitance ($C_g$) of ~$3.6 \times 10^{-19}$ F. This is the normal CB oscillation, corresponding to the gate induced charge number changed by one in the SET island as a function of the gate voltage $V_{G1T}$. However, we can clearly see that, in positive bias region, a fast and short-period oscillation takes place when the gate bias $V_{G1T}$ is scanned above 0.5 V [Fig. 4(a) and 4(b)]. This feature stands in remarkable contrast with the normal gate induced CB oscillation. A further close look at the fast oscillation reveals several distinguished aspects [Fig. 4(b) and 4(c)]. First, the fast oscillation is characterized by rather equally spaced conductance peaks throughout positive biases starting from 0.5 V, with the average spacing value of ~7.4 mV. Second, the spacing between the first peak and the second one is ~11.4 mV, about 1.54 times that of the average value. Third, the evenly spaced short-period oscillation only and uniquely takes place at positive gate biases above 0.5 V. Further, it looks like that the short period oscillation, which possesses fluctuation amplitude comparable to that of the normal SET CB oscillation, is superimposed on top of a slowly varying much larger-period background.

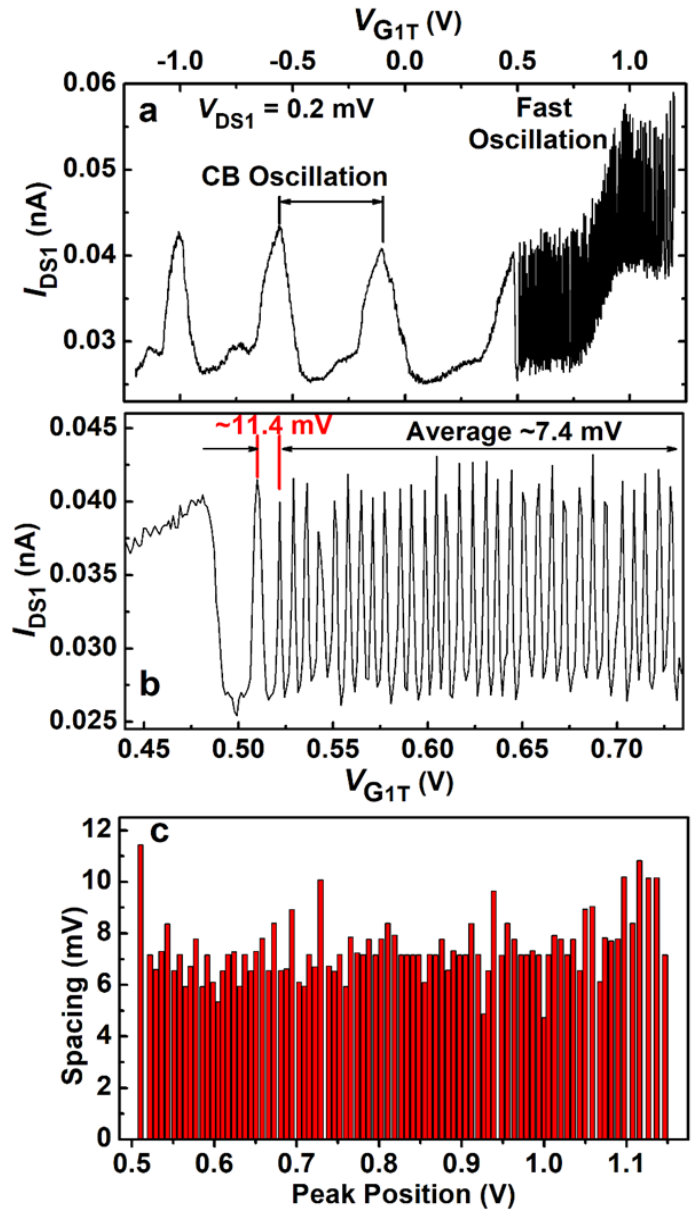

**Fig. 4** Conductance oscillations versus gate bias applied to top gate. (**a**) Current passing through SET1 as a function of the gate bias, $V_{G1T}$, applied to top gate $G_{1T}$. The drain-source bias, $V_{DS1}$, is fixed at 0.2 mV for the measurements. (**b**) Zoomed-in view of the fast oscillation region showing rather uniform and equally spaced resonance peaks with average period of ~7.4 mV. However, the spacing between the 1st and 2nd peaks is ~11.4 mV, about ~1.54 times that of average value. (**c**) The spacing between neighboring peaks versus peak position.

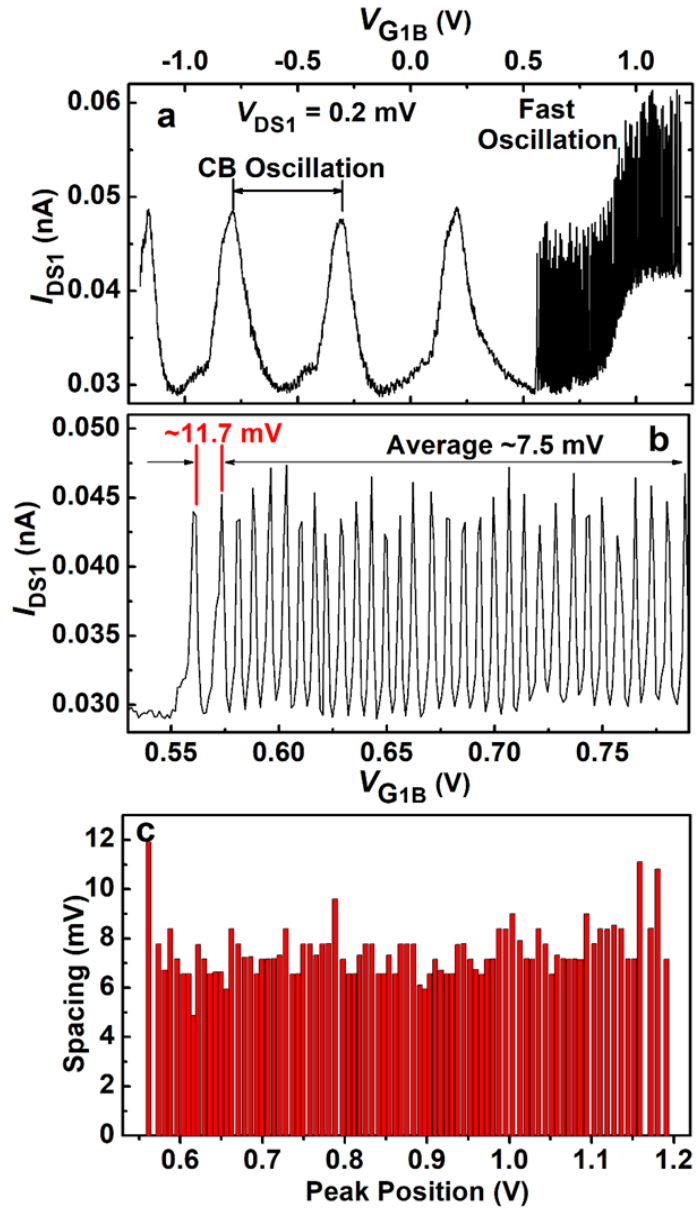

**Fig. 5** Conductance oscillations versus gate bias applied to bottom gate. (**a**) Current through SET1 as a function of the bottom gate bias, $V_{G1B}$, at fixed drain-source bias of 0.2 mV. (**b**) Zoomed-in view of the fast oscillation region showing evenly spaced resonance peaks with average period of ~7.5 mV. The spacing between the 1st and 2nd peaks is ~11.7 mV, about ~1.56 times that of average value. (**c**) The spacing between neighboring peaks versus peak position.

Our experiments produce almost identical output of current signals from SET1 as a function of the gate bias voltage applied to bottom gate $G_{1B}$, as shown in Fig. 5. The SET demonstrates normal CB oscillation with period of ~490 mV in negative bias region [Fig. 5(a)]. When the gate bias $V_{G1B}$ is above 0.55 V, a fast and short-period oscillation appears with average spacing value of ~7.5 mV [Fig. 5(a) and 5(b)]. The spacing between the first resonance peak and the second one is ~11.7 mV, about ~1.56 times that of the average period [Fig. 5(b) and 5(c)]. Measurements on SET2 gave very similar results, demonstrating reproducibility of the observed phenomenon. Further, these experimental observations suggest that the SET has detected the charge state resonances or harmonic motions of charges in a two-level system which is in close proximity to SET [25, 30-33].

Our experimental measurements using SET show a good agreement with the theoretical proposal of the formation of a hydrogen-like Si QHO. The SET fabricated on top of AlGaAs/GaAs heterostructure is parallel-coupled to the QHOs located on δ-doping layers (see Fig. 2). A nearby gate (e.g., gate $G_{1T}$ or $G_{1B}$ in our device configuration) is shared by both of SET and QHO, and can tune and manipulate the quantum states on both. When a negative gate bias voltage is applied and swept, the offset charges on SET island will be adjusted. But the unoccupied energy levels of QHO will shift to higher energy states, as a result of the gate tuning effect. Thus QHO has no effect on SET. Therefore, the SET only demonstrates its normal CB oscillation with larger periodicity as a function of gate induced SET offset charges changing one by one. However, when the applied gate voltages are positive, the QHO discrete unoccupied states shift to lower energies. As the applied positive voltage increases its values, the QHO energy levels keep moving downwards, until one of the unoccupied energy states coincides with $E_f$ of 2DEG. Since the QHO is spatially close to 2DEG, one electron will fill up the originally unoccupied harmonic states of QHO by resonant tunneling from 2DEG. Then the SET, as the ultrasensitive electrometer, gives rise to a resonance peak in the $I$-$V_g$ curve in response to the change of charge states in QHO. This process repeats when QHO confined energy levels of unoccupied states sequentially shift in and out of resonance with $E_f$, starting from the lowest hydrogen-like atomic orbital, which is also degenerate with the harmonic potential minimum, to harmonic energy levels $E_n$ (n = 0,1,2 ….).

We know that, for a three-dimensional (3D) QHO, the energy difference between the potential minimum to the first harmonic energy state is $\frac{3}{2}\hbar\omega_0$, and then evenly spaced harmonic energy levels with energy spacing of $\Delta E = \hbar\omega_0$ take place. For the hydrogenic Si QHO formed in AlGaAs/GaAs heterostructure, its parabolic potential minimum corresponds to the equilibrium ground state when the massive $Si^+$ nucleus locates at the center of electron cloud. This state is originally an unoccupied one since the released valence electron is confined in quantum well. This makes our QHO differing slightly from the normal QHO, where the zero-point energy state (n = 0) is the lowest one. This distinguishing feature is fully reflected in our SET measurements, in which the spacing between the first and second conductance peaks is ~11.5 meV, followed by equally spaced uniform resonance peaks with spacing of ~7.5 meV. Our experimental results revealed by electrical measurements rather quantitatively reproduce our theoretical predictions. Our experiments not only measure the energy levels of the evenly spaced QHO states, also clearly demonstrate that the spacing between the potential minimum to the zero-point energy state ($E_0$) is 1.5 times that of the energy level spacing of the QHO states for an ideal three-dimensional QHO, resulting from the origin of the hydrogenic Si QHO in semiconductor heterostructure and the configuration design of the QHO-SET device architecture. We notice that our experimental value of the harmonic energy spacing is a little smaller than that of theoretical prediction, but the discrepancy is within 15%. We suggest that this discrepancy originates from our simplified theoretical model in which only the general issue of screened Coulomb potential is taken into account which is independent of atom type. A more accurate model should include the "chemical shift" induced by different chemical nature of a specific impurity atom [22-24], and also, the influence from the periodic lattice potential of AlGaAs/GaAs heterostructure.

The electronic states of one or a small number of impurity atoms and defects in semiconductors have long been demonstrated to play a pivotal role in the design and fabrication of some novel devices and/or to generate the otherwise hidden new functionalities in semiconductor devices (see, for example, Refs. 34-36). In particular, Dellow et al. pointed out that the bound states with spatial extent of ~25 nm of a single donor in quantum well directly determine the performance of a gated AlGaAs/GaAs resonant tunneling diode [34]. In this study, we demonstrate that a substitutional Si donor atom will form the hydrogen-like quantum harmonic oscillator (QHO), one of the most fundamental and ubiquitous model systems in the realm of quantum, in an AlGaAs/GaAs heterostructure, which is one of the most widely used semiconductor structures.

## 3  Conclusion

In summary, we have theoretically proposed and experimentally realized the nearly ideal prototype QHO in AlGaAs/GaAs heterostructure. We have implemented the manipulation and detection its QHO states by gate tuning and by utilizing an SET. It is worth noting that the control and operation of QHOs in our device are implemented by a single electrical gate, which has the advantages of reliability, fidelity, and scalability for integration. We expect the generality of our strategy for artificially constructing and high-reliability manipulating QHOs, coupled QHOs, scalable integration in a variety of semiconductor heterostructure materials, such as Si/Ge and other III-V semiconductors. This offers a versatile and well-controlled platform to explore, engineer, and manipulate QHO states at large scales. Using the existing semiconductor materials and fabrication techniques, it is possible to design and construct a series of scalable integration architectures of quantum hardware (e.g., coupled QHOs and QHO-SET) and hybrid hardware (e.g., QHOs integrated with conventional semiconductor transistors). Considering practical functionality and scalability, the integrated QHO systems might offer the prospects for future applications in two emerging fields. The first is terahertz electronics. The energy level spacing of our QHO is about 7.5 meV, corresponding to energy of ~1.8 THz. So the integrated QHO devices might be utilized for THz generation and/or high efficiency detection of electromagnetic radiation above 1.8 THz.

Another possible use of hydrogenic QHO in semiconductor heterostructures is quantum information processing and/or storage. We notice that it is challenging for an QHO to be used as a qubit, as it is difficult to excite and address only two of its states. However, using coupled QHOs with separately tuning gates, we might be able to modify their potential wells to design and construct anharmonic oscillators for qubits.

**Acknowledgements**   We gratefully acknowledge Prof. A. M. Chang for his generous supports in experiments at Duke University, and Dr. F. Altomare for his sincere helps in experiments and careful reading of the manuscript. Fruitful discussions with Profs. Y. Zhang, J. Chen, J. Zhao, and C. Lin are greatly appreciated. M. Feng thanks financial support from the Strategic Priority Research Program of Chinese Academy of Sciences (Grant No. XD30000000), and from the National Natural Science Foundation of China (Grant Nos. 11574364 and 11774267). L. Mao thanks financial support from the National Key R&D Program of China by the Ministry of Science and Technology of China (Grant No. 2015C8932400).